\documentclass[twoside,12pt]{article}
\usepackage{epsf,epsfig,amssymb,amsmath,graphicx,float}
\usepackage{color}
\usepackage{cite}
\usepackage{amsmath}
\usepackage{hyperref}
\usepackage{indentfirst}
\begin{document}
\renewcommand{\thefootnote}{\fnsymbol{footnote}}

\begin{titlepage}

\begin{center}

\vspace{1cm}

{\Large {\bf Non-standard cosmological scenarios, Sommerfeld enhancement and 
asymmetric dark matter}}

\vspace{1cm}

{\bf Reyima Rashidin, Fangyu Liu\footnote{liufgyu@hotmail.com (Corresponding 
author)}, Hoernisa Iminniyaz\footnote{wrns@xju.edu.cn (Corresponding author)}}

\vskip 0.15in
{\it
{School of Physics Science and Technology, Xinjiang University, \\
Urumqi 830017, China} \\
}

\abstract{We discuss the relic density of asymmetric dark matter with
  long-range interactions in non-standard cosmological scenarios where the extra
  cosmic energy is introduced. The Hubble expansion rate is modified
  in non-standard cosmological models, which affects the relic density of dark
  matter. If the mass of the dark exchanged gauge boson is less than the mass
  of dark matter particles multiplied by the dark fine-structure constant, the
  wave function of the incoming particles of dark matter annihilation would be
  distorted away from the free plane-wave approximation, yielding the
  significant enhancements to annihilation cross sections, known as Sommerfeld
  enhancement. Sommerfeld enhancement results the asymmetric dark matter relic
  density under abundant, while the enhanced cosmic expansion rate increases the
  relic density of asymmetric dark matter by letting the decoupling from the thermal equilibrium occurs earlier. However, the mixed effect of modified cosmic expansion and the Sommerfeld 
 enhancement on asymmetric dark matter density
  evolution  becomes quite complicated, due to their opposing affects on relic density and contrary strength developments as the temperature fell. We investigated the mixed effects on
  asymmetric dark matter freeze-out process and current relic abundance, which
  is significantly different from the standard situation. Further, we 
  calculate the constraint relations of the asymmetric dark matter 
  pertubative annihilation
  cross section and dark coupling constant with the mass of asymmetric dark
  matter, the asymmetry factor with the dark coupling constant, when the relic
  density of asymmetric dark matter falled in the observed region. The upper
  bounds on the mass of asymmetric dark matter for s-wave and p-wave
  annihilations are also derived. 
}
\end{center}
\end{titlepage}
\setcounter{footnote}{0}

\section{Introduction}

Asymmetric Dark Matter (DM) scenarios have been put forth for explaining why
the average density of baryons is comparable to that of DM, $\Omega_{\rm DM}
\simeq 5 \Omega_b$\cite{Planck:2018vyg}. For asymmetric DM, particles 
and 
anti-particles 
are not self-conjugate, and the final DM relic density is determined by the 
asymmetry between the DM particle and anti-particle. The idea of asymmetric 
DM provides the possible connection between the baryon asymmetry and 
postulated DM asymmetry\cite{Hooper:2004dc,Nardi:2008ix,An:2009vq,Iminniyaz:2011yp,Cohen:2009fz,Kaplan:2009ag,Cohen:2010kn,Shelton:2010ta,Kang:2011cni,
Ellwanger:2012yg,Petraki:2013wwa,Boucenna:2013wba,Zurek:2013wia,Mahapatra:2023dbr,Borah:2024wos}. 

It is natural that DM has new interactions with the standard model 
particles or among themselves from a theoretical point of view. If these new
interactions are taken into account, macroscopic quantities such as DM relic
density or DM distributions in galaxies would be different. In the context of
DM models, a new light exchanged force-carrying boson, together with the
non-perturbative effect it induced, has been studied continuously over the
recent two
decades\cite{Baldes:2017gzw,Bollig:2024ipe,Arkani-Hamed:2008hhe,Sommerfeld:1931qaf,Qiu:2024iyo,Iengo:2009ni,Slatyer:2009vg,
Petraki:2015hla,vonHarling:2014kha,Petraki:2016cnz,Petraki:2014uza,Kamada:2020buc,Cassel:2009wt,Duerr:2018mbd,Feng:2010zp,Cirelli:2016rnw,Feng:2009mn,Agrawal:2017rvu,Agrawal:2016quu,Chen:2023rrl}. According
to energy-time uncertainty principle, the range of force of the exchanged
boson is approximately inversely proportional to its mass. When the range of
mediated force $\propto {m_{med}}^{-1}$ is larger than the Bohr radius of DM
particle pair $\propto (\alpha m)^{-1}$, where $\alpha$ is dark fine-structure
constant, this case is called long-ranged potential. The long-ranged effect
causes the incoming particle's wave function of annihilation process cannot be
considered as free state anymore at the origin. Therefore, the cross section
can not be fully calculated in the pertubative methods and the cross section
needs to be corrected. This effect is known as Sommerfeld effect. 
If the mediator is massless, the Sommerfeld effect is naturally present as the
condition is fulfilled. Whereas for the massive mediator, the effect depends
on the mass ratio between the mediator and particle pairs.  In some cases, in
order to calculate the cross section, the above process can be decomposed into
two separable parts that the high energy (hard) and low energy (soft) momentum
transfer process. The cross section of hard annihilation process is calculated
with typical perturbative methods. The Sommerfeld correction is described as a
re-summation of soft mediator exchanges and solely depend on the potential
generated by mediator. Under the assumption that the Sommerfeld factor is
separable from the hard matrix element, the annihilation cross section is
written as $\sigma_{ann}v_{rel}=\sum\limits_{l=0}^{\infty}
\sigma_lv_{rel}^{2l}S_{ann,l}$, where $S_{ann,l}$ is the Sommerfeld factor
calculated by utilizing the scattering wave functions of the corresponding
$\rm Schr\ddot{o}dinger$ equation, $l$ indicates the expansion in 
partial waves, when
$l=0,1$, $S_{ann,0},\  S_{ann,1}$ are s-wave and p-wave Sommerfeld factors respectively. As a result, the annihilation rate is enhanced at low velocities . Sommerfeld effect suppresses the
couplings needed to achieve the observed DM density through the freeze-out in
the early universe. On the contrary, the late-time DM annihilation signals are
enhanced due to the Sommerfeld effect.

The relic density of asymmetric DM is determined by the thermal freeze-out
mechanism which assumes the asymmetric DM particles were in thermal
equilibrium in the early universe and decoupled from equilibrium when the
universe is radiation dominated. The relic density is obtained by solving the
Boltzmann equation which describes the evolution of number density of the
particles in the expanding universe. In the Boltzmann equation, the Hubble
expansion rate plays an extreme significant role, and in the following we
would see that the Boltzmann equation in various typical non-standard
cosmological models differs only in the rate of cosmic expansion. Standard 
cosmological scenario assumes the universe is radiation dominated when DM 
freezes out.

The time when DM began to freeze out was incredibly early, far earlier than $1s$ after Big Bang, 
which is far beyond the reach of observations. No one could resist their curiosity and refrain from exploring some different cosmological scenarios. Indeed, in the recent two decades, the affect of non-standard cosmological scenarios on the DM density has been investigated continuously \cite{Joyce:1996cp,Iminniyaz:2016iom,Drees:2007kk,Drees:2006vh,Salati:2002md,BINETRUY2000285,Iminniyaz:2018das,Binetruy:1999ut,DEramo:2017gpl,Langlois:2002bb,Guo:2009nt,Cicoli:2023opf,Catena:2009tm,Hertzberg:2024uqy,Gron:2024vmf,Okada:2004nc,AbouElDahab:2006glf,Schelke:2006eg,Giare:2024akf,Chanda:2021tzi}.
The non-standard scenarios, such as the kination model, the brane world
cosmology predict faster expansion rate of the universe. In kination model the 
kinetic energy of scalar field dominated over the early era, in brane world
cosmology, the extra dimension of spacetime changed the Friedmann equation. The modification of the cosmic expansion rate leaves its imprint on the asymmetric DM relic density.  

In our previous work\cite{Qiu:2024iyo}, we discussed the impact of Sommerfeld enhancement on the
relic density of asymmetric DM in the standard cosmological
scenarios. In the present work, we investigate the effect of Sommerfeld enhancement with massive gauge boson on the relic density of asymmetric DM in non-standard cosmological scenarios.
The Sommerfeld effect lonely would decrease the relic density of DM because of
the boosted annihilation cross section. On the other hand, the enhanced expansion rate of universe would increase the relic density of asymmetric DM because of the earlier decoupling point. However, the mixed effect of Sommerfeld enhancement and modification 
of cosmic expansion rate is quite complicated, in which the mixed effect is not the simply combined operation. The density evolution process is quite different from the standard case. Indeed, as the hypotheses about DM density evolution become increasingly complex, it is difficult to simply yet clearly describe the density evolution with using few phenomenological quantity that have well visual and obvious physical interpretation like the standard case.
In our work, we try to find the common parameter space
which can fulfill the observed relic density and the coupling strength demanded for cold DM. We would consider two minimal scenarios that the asymmetric DM
couples to the light vector mediator or the scalar mediator. 

The paper is arranged as following. Boltzmann equation and the cosmic
expansion rate in non-standard cosmological models are reviewed in the next
section. In the third section, we calculate the relic density of asymmetric DM
with Sommerfeld enhancement in non-standard cosmological scenarios.  In the
fourth section, we present the density evolution of asymmetric DM and analyze the mixed
effect of Sommerfeld enhancement and the modification of cosmic expansion on
asymmetric DM density evolution. Then we derive the constraint relations
of the annihilation cross section and coupling constant with the mass of 
asymmetric DM, the asymmetry factor with the coupling constant. 
    
\section{Boltzmann equation and the cosmic expansion rate in non-standard cosmological models }
Following we shortly review the Boltzmann equation and the expansion rate of universe\cite{CBLiang,reviewBE,Weinberg:2008zzc,Kolb:1988aj}.
The Boltzmann equation that describes the
evolution of DM number density, is
\begin{equation}\label{fBoltzmann}
L[f]=C[f]\,,
\end{equation}
where $f(x^\mu,p^\mu)$ is the phase space distribution function of particle number, $C[f]$ is the collision term which is not generally influenced by cosmology. $L[f]$ is the covariant Liouville operator, 
\begin{equation} \label{Liouville}
L=p^\mu\partial_\mu-{\Gamma^\mu}_{\sigma \nu}p^\sigma p^\nu\frac{\partial}{\partial p^\mu}\,,
\end{equation}
where $\Gamma$ is Christoffel symbol. Indeed, Eq.\eqref{Liouville}
can be obtained by expanding $df(x^\mu,p^\mu)/d(\tau/m)$ and
substituting it into
geodesic equation 
${\rm d}^2x^\mu/{\rm d}\tau^2+{\Gamma^\mu}_{\sigma\nu}({\rm d}x^\sigma/{\rm
  d}\tau)({\rm d}x^\nu/{\rm d}\tau)=0$. For the
Robertson-Walker metric
${\rm d}s^2=-{\rm d}t^2+a^2(t)[{\rm d}r^2/(1-kr^2)+r^2({\rm
    d}\theta^2+sin^2\theta {\rm d}\varphi^2)]$,
assuming $f$ is spatially homogeneous and isotropic, the Liouville term becomes 
\begin{equation}\label{RWLiouville}
L[f(E,t)]=E\frac{\partial f}{\partial t}-H{\lvert \vec{p}\rvert}^2\frac{\partial f}{\partial E}\,.
\end{equation}
For non-standard cosmological models, the Liouville term may be changed, such
as the Bianchi type I models which assumes the universe is homogeneous but
anisotropic before Big Bang Nucleosynthesis (BBN). In this
article, the non-standard cosmological models considered is adding an extra
energy content into the Friedmann-Robertson-Walker models. 
Usually, additional energy fields are often introduced to solve certain cosmological problems, such as the dark energy field which is added to solve the cosmic accelerated expansion mechanism, or
the scalar field that causes the inflation that is included in order to solve 
the horizon problem. 
Without changing the metric of spacetime, adding an extra energy term does not change the abstract form of the Liouville term and the Boltzmann
Equation. However, the concrete form of expansion rate of the universe $H$
would be changed due to the additional energy. In other words, the change of
$H$ that appeared in the Boltzmann equation, would cause the difference in the
evolution of particle number densities. 

We introduce the extra energy density $\rho_D$ as 
\begin{equation}\label{rhoD}
\rho_D=\rho_D(T_r)\left(\frac{T}{T_r}\right)^{n_D}\,,
\end{equation}
where $n_D$ is the constant which parametrizes the behavior of the energy
density, $T_r$ is reference temperature which we subsequently take the value
as $T_r=4\,$GeV. The form of energy in Eq.(\ref{rhoD}) is consistent with the 
kination model physically and mathematically when $n_D=6$, and matches with 
the brane world model mathematically when $n_D=8$, but not physically, in that 
model the extra dimension of space-time changes the Friedmann equation.

Based on the analysis above, we discuss the affect of extra energy density 
on the cosmic expansion
rate and the relic density of DM. The relationship between the expansion rate and energy is described by the Friedmann equation which is the $0-0$ component of Einstein field equation,
\begin{equation}
H_D^2=\frac{8\pi G}{3}(\rho_{\rm rad}+\rho_D)\,,
\end{equation}
where $\rho_{\rm rad}$ is the radiation energy density. 
Here, we are concerned with the period of the very early universe that the
radiation energy is dominated over the matter energy and the term 
containing the constant $k$ that represents three kinds of space of constant
curvature can be neglected. For convenience, we 
define $\eta\equiv
\rho_D(T_r)/\rho_{rad}(T_r)$, then the cosmic expansion rate can be expressed by
\begin{equation}\label{H_D}
H_D^2=\frac{8\pi G}{3}\rho_{\rm rad} \left[1+\eta\frac{g_*(T_r)}{g_*(T)}\left(\frac{T}{T_r}\right)^{n_D-4}\right]\,.
\end{equation}

Integrating Eq.\eqref{fBoltzmann}, the Boltzmann equation in non-standard 
cosmological model becomes
\begin{equation}\label{BoltzmannN}
\frac{{\rm d}n_{\chi(\bar{\chi})}}{{\rm d}t}+3H_D n_{\chi(\bar{\chi})}=
-\langle \sigma v \rangle  (n_\chi n_{\bar{\chi}}
-n_{\chi,{\rm eq}}n_{\bar{\chi},{\rm eq}})\,,
\end{equation}
where $n_{\chi(\bar{\chi})}$ is the number density of DM particle or
anti-particle, $\langle \sigma v \rangle $ is the thermal average of
annihilation cross section times relative velocity of annihilating asymmetric 
DM particles and anti-particles.

Simultaneously, in order not to conflict with other theories, the modification 
of cosmology we considered, are not arbitrary. We considered the BBN constraints
which is usually the first significant event that occurs after the DM freeze
out. The modification of expansion rate occurs only before BBN and returns
to the standard case at the beginning of BBN. The BBN constraint can be 
taken simply as following  
\begin{equation}\label{BBNlimit}
\rho_D ({1\rm MeV})\,\textless \, \rho_{\rm rad}({\rm 1MeV}) \quad i.e. \quad
\eta\textless \left(\frac{Tr}{{\rm 1MeV}}\right)^{n_D-4}\,.
\end{equation}
Eq.\eqref{BBNlimit} ensures that radiation dominated era returns when the 
temperature of the universe is 1 MeV. If choosing
$n_D\textgreater 4$ and using Eq.\eqref{BBNlimit}, 
$\rho_D$ (Eq.\ref{rhoD}) will always be less significant at any time after 
the BBN. 

\section{Sommerfeld enhancement and the relic density of asymmetric DM}
In this section, we calculate the relic density of asymmetric DM with
long-range interactions in non-standard cosmological scenarios. A convenient
and useful method of solving the Boltzmann equation \eqref{BoltzmannN} is to introduce $Y_{\chi(\bar\chi)} =n_{\chi(\bar\chi)}/s$ and $x = m/T$\cite{Kolb:1988aj}. Here 
$s= 2 \pi^2 g_{*s}/45\, T^3$ is the entropy density, and $g_{*s}$ is the 
effective number of entropic degrees of freedom. Assuming the entropy is conserved per comoving volume, then Eq.\eqref{BoltzmannN} can be expressed as 

\begin{equation} \label{eq:boltzmann_Y}
\frac{{\rm d} Y_{\chi(\bar\chi)}}{{\rm d}x} =
      - \frac{H_{\rm std}}{H_D}\, \frac{\lambda \langle \sigma v \rangle}{x^2}\,
     (Y_{\chi}~ Y_{\bar\chi} - Y_{\chi, {\rm eq}}~Y_{\bar\chi, {\rm eq}}   )\,,
\end{equation}
where
$\lambda = 1.32\,m M_{\rm Pl}\, \sqrt{g_*}\, $, $g_*\simeq g_{*s}$ and
$dg_{*s}/dx\simeq 0$, and $H_{\rm std}$ is the cosmic expansion rate in the
standard cosmology. Here $M_{\rm Pl} =2.4 \times 10^{18}$ GeV is the reduced 
Planck mass.
By subtracting the Boltzmann equations for $\chi$ and $\bar\chi$ in 
Eq.(\ref{eq:boltzmann_Y}), we obtain $Y_{\chi} - Y_{\bar\chi} = C\,$, where 
$C$ is a constant. Finally, we rewrite the Boltzmann equation 
\eqref{eq:boltzmann_Y} as
\begin{equation} \label{eq:Yeta}
\frac{{\rm d} Y_{\chi(\bar\chi)}}{{\rm d}x} =
     -\frac{H_{\rm std}}{H_D}\,  \frac{\lambda \langle \sigma v \rangle_s}{x^2}~  
     (Y_{\chi(\bar\chi)}^2 \mp C Y_{\chi(\bar\chi)}  - Y^2_{\rm eq})\,.
\end{equation}
where
\begin{equation}
\frac{H_D}{H_{\rm std}}=\sqrt{1+\eta\left(\frac{m}{x\, T_r}\right)^{n_D-4}}\,.
\end{equation}
Here the annihilation cross section is substituted by $\langle  \sigma v
\rangle_s$ which is thermally averaged Sommerfeld enhanced
annihilation cross section  
$\langle  \sigma v \rangle_s  =  a\, \langle S_s \rangle +  b\, \langle v^2\,S_p \rangle\,$.
For the massive mediator, the Sommerfeld enhancement factors are\cite{Kamada:2020buc,Cassel:2009wt,Feng:2010zp,Duerr:2018mbd}
\begin{eqnarray}
&&S_s = \frac{2\pi \alpha}{v} 
   \frac{{\rm sinh}\left(\frac{6 m v}{\pi m_{\phi} }\right)}
   {{\rm cosh}\left(\frac{6 m v}{\pi m_{\phi} }\right) - 
    {\rm cos}\left(2\pi\sqrt{\frac{6 \alpha m}{\pi^2 m_{\phi}} - 
      \frac{ 9 m^2 v^2}{\pi^4 m^2_{\phi} }}\right)}\qquad {\rm s-wave\;annihilation}\\		
			&&S_p = \frac{\left(\frac{6 \alpha m}{\pi^2 m_{\phi}} -1 \right)^2 + 
           \frac{36 m^2 v^2}{\pi^4 m_{\phi}^2}} 
            { 1+ \frac{36 m^2 v^2}{\pi^4 m_{\phi}^2}}  \, S_s
            \,\,\,\,\,\,\,\,\,\,\,\,\,\,
\qquad \qquad \qquad \qquad {\rm p-wave\; annihilation}
\end{eqnarray}
where $v$ is the relative velocity of the
annihilating asymmetric DM particle $\chi$ and anti-particle $\bar\chi$. 

We would consider two minimal cases in which the asymmetric DM couples to a light massive vector or scalar boson. One is $\chi\bar\chi \rightarrow 2 \gamma$,
asymmetric DM annihilating into two dark photons. 
Assuming that DM is Dirac Fermions and couples to the dark photon through local gauged $U(1)$ interaction, formally just as ordinary Dirac particles coupling to photons, but with a dark coupling strength. Then, the Lagrangian is written as $\mathcal{L}=\mathcal{L}_{\rm Dirac}+\mathcal{L}_{\rm Maxwell}+\mathcal{L}_{\rm int}=\bar{\psi}(i\gamma^\mu\partial_\mu-m_{DM})\psi-{F_D}_{\mu\nu}{F_D}^{\mu\nu}/4-g_{D}\bar{\psi}\gamma^{\mu}\psi{A_D}_{\mu}$, where subscript $D$ represent $"$dark$"$ quantity. In this case, it is an s-wave process and its Sommerfeld enhanced annihilation cross section times relative velocity is\cite{Baldes:2017gzw,Petraki:2015hla,vonHarling:2014kha,Petraki:2016cnz}  
\begin{equation}
    (\sigma v )_{\rm s(s-wave)} = \sigma_0\,S_s=\frac{\pi \alpha^2}{m^2}\,S_s\,,
\end{equation}
where the leading order cross section times the relative velocity $\sigma_0$ is perturbative value in the non-relativistic regime, $\alpha=g_D^2/(4\pi)$ is dark fine-structure constant. Another case we considered is $\chi \bar\chi \rightarrow 2 \varphi $, asymmetric DM
annihilating into two scalars. In this case, DM is assumed to be coupled to scalar particle through Yukawa interaction, and  the Lagrangian is $\mathcal{L}=\mathcal{L}_{\rm Dirac}+\mathcal{L}_{\rm Klein-Gordon}+\mathcal{L}_{\rm int}=\bar{\psi}(i\gamma^\mu\partial_\mu-m_{DM})\psi+\partial_\mu\phi\partial^\mu\phi/2-m^2\phi^2/2-g_D\bar{\psi}\psi\phi$. 
It is p-wave process and its Sommerfeld enhanced cross section is   
\begin{equation}
    ( \sigma v )_{\rm s(p-wave)} = \sigma_1 \,v^2 S_p=\frac{3 \pi
    \alpha^2}{8 m^2}\,v^2 S_p \,,
\end{equation}
where $\sigma_1$ is the tree level cross section. 

To confirm the initial value of the Boltzmann equation, we follow the standard
picture of DM particle evolution\cite{Scherrer:1985zt}. In the early universe prior to BBN,
the asymmetric DM particles and anti-particles were assumed in thermal
equilibrium. When the temperature falls, the particle interaction rate drops
to less than the cosmic expansion rate, the asymmetric DM decoupled from
thermal equilibrium and subsequently formed the DM relic in the current universe. Therefore, the initial value can adopt the Maxwell-Boltzmann thermal equilibrium statistics, i.e.
$n_{\chi(\bar\chi),{\rm eq}} = g_\chi 
{\big[m T/(2 \pi) \big]}^{3/2}{\rm e}^{(-m \pm  \mu_\chi)/T}$. The value of chemical potential would be a question.
For solving the relic density of asymmetric DM, a useful approach which finely 
addressed the chemical potential was proposed in Ref.\cite{Iminniyaz:2011yp}.
Closely following the method which is used in Ref.\cite{Iminniyaz:2011yp},  
the total final DM relic density is expressed as  
\begin{eqnarray} \label{eq:omega}
 \Omega_{\rm DM}  h^2 & = & 2.76 \times 10^8\, \left[ Y_{\chi}(x_\infty) + 
                       Y_{\bar\chi}(x_\infty) \right]\,m,
\end{eqnarray}
where $\Omega_{\chi} = \rho_{\chi}/\rho_c$ with 
$\rho_{\chi}=n_{\chi} m = s_0 Y_{\chi}  $ and 
$\rho_c = 3 H^2_0 M^2_{\rm Pl}$, here $s_0 \simeq 2900$ cm$^{-3}$ is the 
present entropy density, and $H_0$ is the Hubble constant. 

\section{Effect of Sommerfeld enhancement and Constraints}

\begin{figure}[h]
  \begin{center}
     \hspace*{-0.5cm} \includegraphics*[width=8cm]{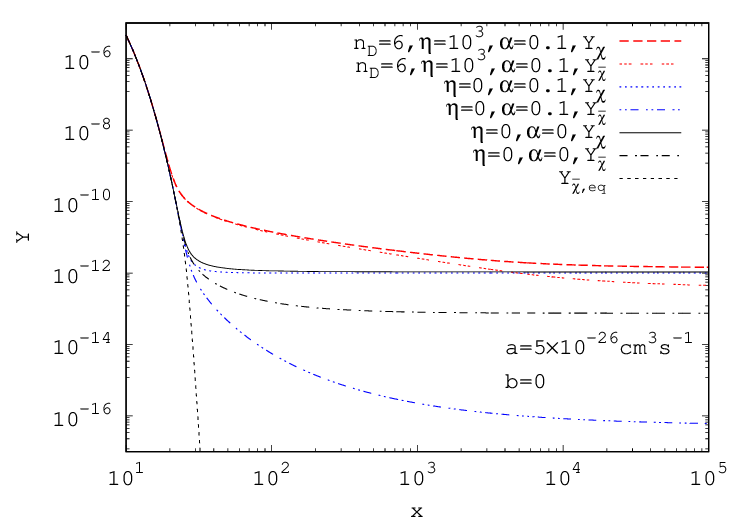}
    \put(-115,-12){(a)}
    	\vspace{0cm}
    \hspace*{-0.5cm} \includegraphics*[width=8cm]{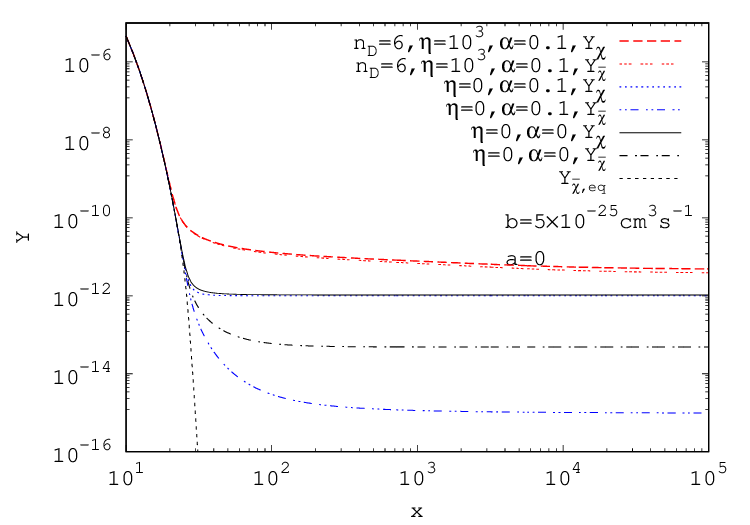}
    \put(-115,-12){(b)}
    	\vspace{0cm}
         \hspace*{-0.5cm} \includegraphics*[width=8cm]{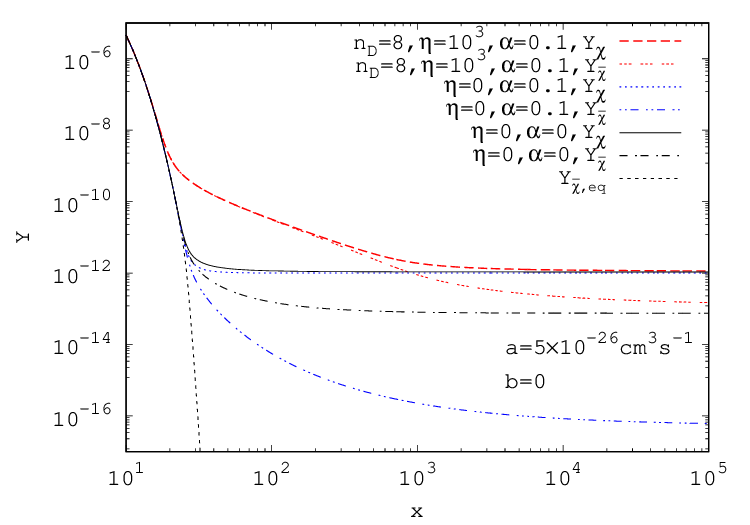}
    \put(-115,-12){(c)}
    	\vspace{0cm}
    \hspace*{-0.5cm} \includegraphics*[width=8cm]{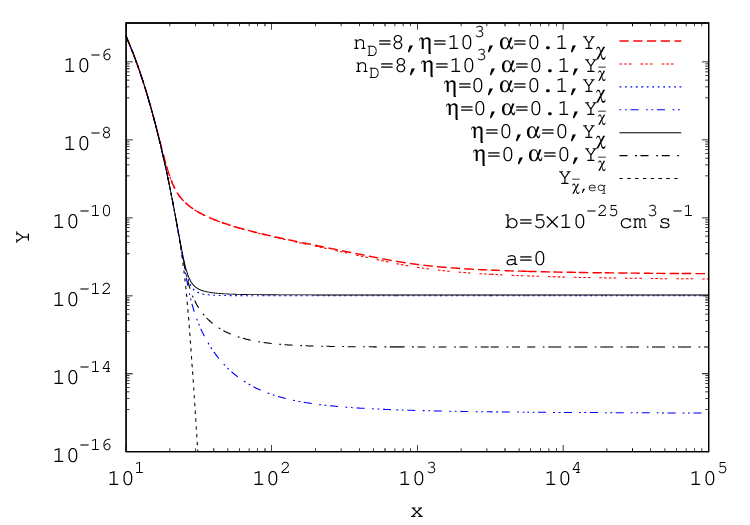}
    \put(-115,-12){(d)}
     \caption{\label{yx} \footnotesize
  Asymmetric DM abundances $Y_{\chi}$ and $Y_{\bar\chi}$ as a function of
  $x$. Here $m = 500$ GeV, $m_{\phi} = 0.25$ GeV, $g_{\chi} = 2$, $g_* = 90$.}  
      \end{center}
\end{figure}
In Fig.\ref{yx}, we plot the evolution of asymmetric DM abundances of
$Y_{\chi}$ and $Y_{\bar\chi}$ as a function of inverse-scaled temperature
$x$. Here $\eta=10^3$, $\alpha=0.1$ and $n_D=6$ or $n_D=8$ correspond to the asymmetric DM
abundances $Y_{\chi}$ (dashed red line) and $Y_{\bar\chi}$ (long dashed
red line) in the non-standard cosmological scenarios with Sommerfeld effect, 
respectively. $\eta=0$ and $\alpha=0.1$ (dotted blue line and double 
dot-dashed blue line) are for the standard cosmological
scenarios including Sommerfeld enhancement while $\eta=0$ and $\alpha = 0$
(solid black line and dot-dashed black line)
returns to the situation when there is no Sommerfeld enhancement. We 
noted that both the asymmetric DM particle abundance $Y_{\chi}$ and anti-particle
abundance $Y_{\bar\chi}$ are affected by Sommerfeld enhancement. The effect on
anti-particle's abundance is more visible. Compared the dot-dashed black 
line $Y_{\bar\chi}(\eta=0,\alpha=0)$ with the double dot-dashed blue line 
$Y_{\bar\chi}(\eta=0,\alpha=0.1)$ in panel $(a)$, we can see the relic
abundance of anti-particle is decreased from $7.51\times10^{-14}$ to
$6.13\times10^{-17}$ in the standard cosmological scenario, which is three 
orders of magnitude. Making an analogy between the 
dot-dashed black line with the double dot-dashed
blue line, we find the double dot-dashed blue line does 
not flatten quickly. Morever, it 
continues to decrease slowly. That is because of the enhanced annihilation cross
section due to the Sommerfeld effect as the
temperature drops. The Sommerfeld factor is increased following the decrease
of temperature, for example, $\langle S_s \rangle(x=10)= 1.70$, 
$\langle S_s
\rangle(x=10^2)= 3.73$, $\langle S_s \rangle(x=10^3)= 11.22$, $\langle S_s
\rangle(x=10^4)= 35.48$, $\langle S_s \rangle(x=10^5)= 113.17$ in Fig.\ref{yx}. 
We can see that, around the typical decoupling temperature
$(x\sim 20)$, the effect of Sommerfeld enhancement is not significant, which
accounts for the point that double-dot dashed blue line deviates from
$Y_{\bar{\chi},eq}$ is nearly the same as the point that dot-dashed black
line deviates from $Y_{\bar{\chi},eq}$. 

On the other hand, for the non-standard cosmological scenarios with Sommerfeld 
effect, $Y_{\chi}$ and $Y_{\bar\chi}$ are both increased, i.e, in panel $(a)$
of Fig.\ref{yx},  the abundance of anti-particle 
$Y_{\bar\chi}(\eta=10^3,\alpha=0.1)$ 
(long dashed red line) is rather sensitive to 
the enhanced expansion rate. 
The long dashed red line is evidently higher than the double dot-dashed blue 
line $Y_{\bar\chi}(\eta=0,\alpha=0.1)$. For 
example in panel $(a)$, the abundance of anti-particle 
$Y_{\bar\chi}(\eta=10^3,\alpha=0.1)$ is increased from
$6.13\times10^{-17}$ to $4.51\times10^{-13}$, which is caused by the enhanced
cosmic expansion rate. We can see, the point at 
$Y_{\bar{\chi}}(\eta=10^3,\alpha=0.1)$ deviates from
$Y_{\bar\chi,{\rm eq}}$ is earlier than the point that dot-dashed black or
double dot-dashed blue lines. The reason is following, 
the enhancement factor of cosmic expansion rate $H_D/H_{\rm std}$ is quite large 
around $x\sim 10$ as $H_D(x=10)/H_{\rm std}(x=10)= 395.29$ in panel $(a)$, while 
the enhancement factor of Sommerfeld effect $\langle S_s \rangle$ is not
visible as $\langle S_s \rangle(x=10)= 1.70$. The additional energy term
enhanced the expansion rate of the universe, and then the cosmic expansion
rate at an earlier moment already exceeds the Sommerfeld enhanced particle interaction rate. 
 That also caused, for a not-long period of time, the dashed red line and
 long dashed red lines around the decoupling temperature are remarkably higher 
than the solid black and dotted blue lines.
As temperature falls, the cosmic expansion enhancement factor is continuously 
decreased to less than $1$ before BBN era 
($\rho_D(T=1{\rm MeV})/\rho_{\rm rad}(T=1{\rm MeV})\textless 1$). In contrast, 
Sommerfeld enhancement factor is continuously increased. As a result, the 
previous comparative condition is unable to persist.
Therefore, over relatively long period of time, the dashed red line gradually
approaches the black solid line and in the end the difference between them 
becomes very small. In Fig.\ref{yx}, we can see that, in the presence of the
Sommerfeld effect, the current relic density of DM particles $Y_{\chi}$ in
non-standard cosmology where the cosmic expansion rate is enhanced, is close
to that in standard cosmology in panels (a) and (c), while this is not the
case in panel (b) and (d). That is to say, if the enhancement factor of cosmic
expansion is not sufficiently large, then even though the point of decoupling
from thermal equilibrium value is advanced, the final relic abundance of DM
particles $Y_{\chi}(\eta=10^3,\alpha=0.1)$ , due to the gradually increased
Sommerfeld factor as the temperature fell, becomes close to that of standard
cosmology $Y_{\chi}(\eta=0,\alpha=0.1)$. On the contrary, if the enhancement of cosmic expansion is taken much larger or the enhancement of Sommerfeld effect is taken much smaller, then the difference between $Y_{\chi}(\eta,\alpha)$ and $Y_{\chi}(\eta=0,\alpha)$ will be considerable.

% Fig. 2
\begin{figure}[h]
  \begin{center}
     \hspace*{-0.5cm} \includegraphics*[width=8cm]{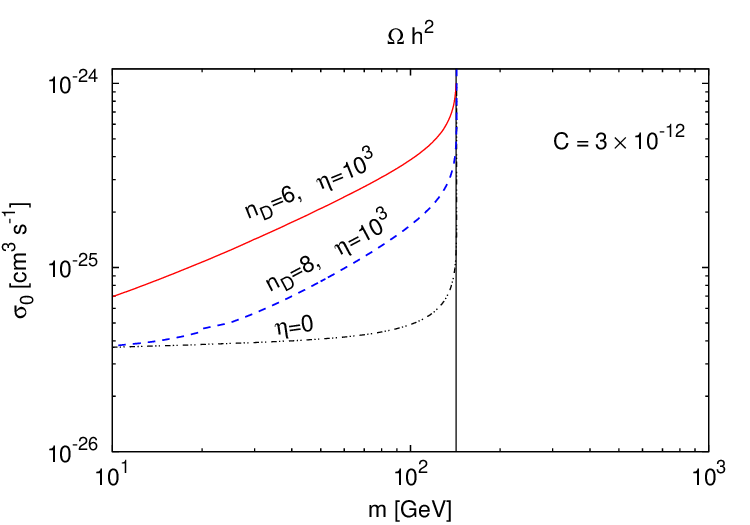}
    \put(-115,-12){(a)}
    \vspace{0cm}
    \hspace*{-0.5cm} \includegraphics*[width=8cm]{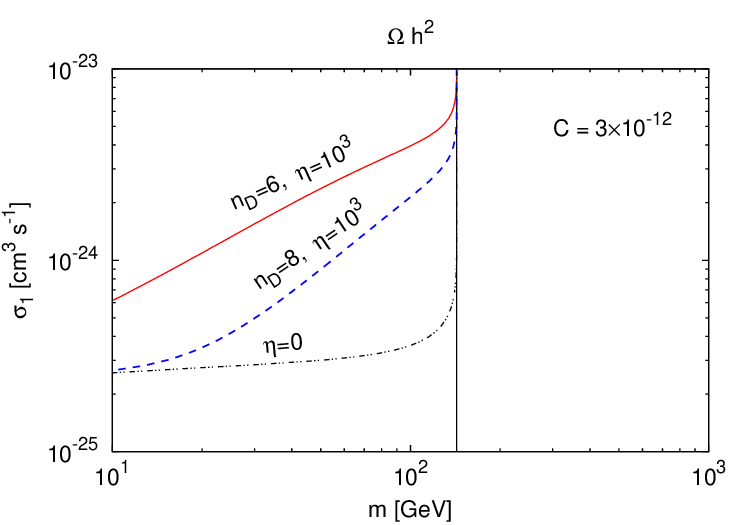}
    \put(-115,-12){(b)}
        	\vspace{0cm}
         \hspace*{-0.5cm} \includegraphics*[width=8cm]{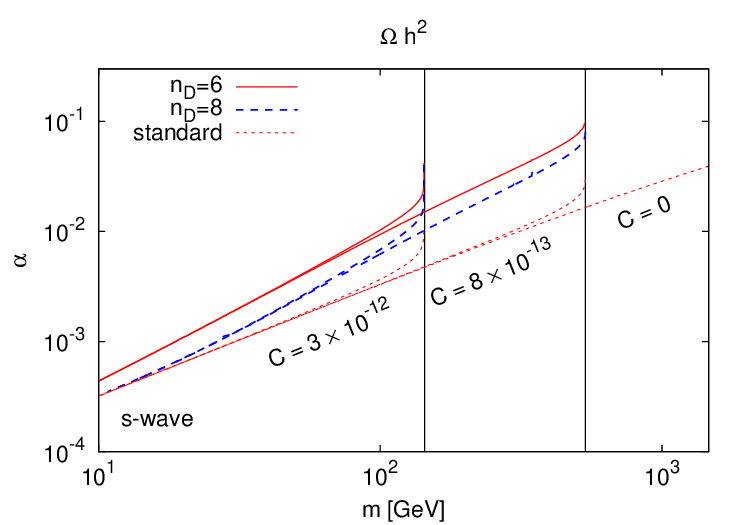}
    \put(-115,-12){(c)}
    	\vspace{0cm}
    \hspace*{-0.5cm} \includegraphics*[width=8cm]{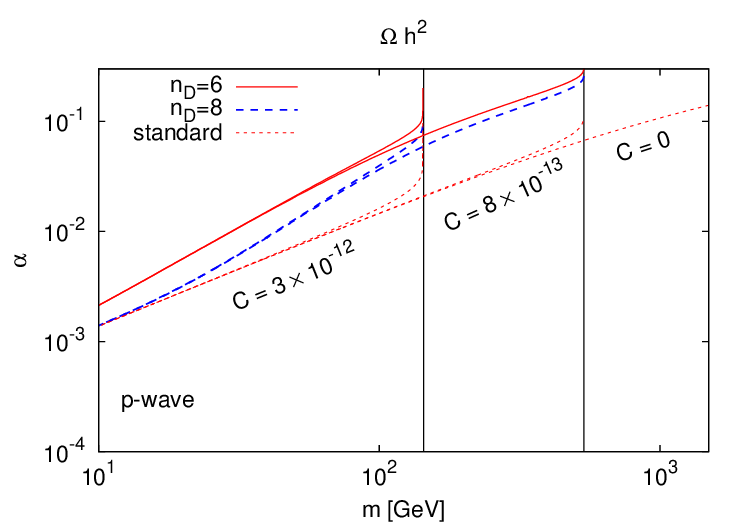}
    \put(-115,-12){(d)}
     \caption{\label{sigma-m} \footnotesize
Contour plots of the perturbative annihilation cross section $\sigma_0$
($\sigma_1$) and the coupling constant $\alpha$ with mass $m$, when 
$\Omega_{DM}h^2=0.120$.
 Here $m_{\phi} = 0.25$ GeV, $g_{\chi} = 2$, $g_* = 90$.}  
      \end{center}
\end{figure}

In Fig.\ref{sigma-m}, we show the relations of the 
perturbative annihilation cross section parameters $\sigma_0(\sigma_1)$ and 
the dark fine-structure constant $\alpha$ with mass $m$ for the observed
value of DM abundance $\Omega_{DM} h^2=0.120$. From panels $(a)$ and $(b)$, we 
can see that, for a certain
mass $m$, $\sigma_0(\sigma_1)$ is increased in the non-standard cosmological
scenarios where the expansion rate is enhanced. The lines corresponded to 
$n_D=6$ is higher than the lines with $n_D=8$. If the point that asymmetric DM
decoupled from equilibrium $Y_{\bar{\chi},eq}$ is later than $T_r$, which is
indeed typical assumption, then for the same value of $\eta$, the smaller $n_D$
corresponds to the larger values of $\sigma_0(\sigma_1)$. If the decoupling 
point is earlier than $T_r$,
the case would be inverse. We also see the enhanced cosmic expansion rate does
not change the maximum value of asymmetric DM mass $m$.
As $\eta$ is gradually decreased to $0$, the situation would return to the
standard case. The broadened range of parameter spaces of DM annihilation
cross sections comes at the cost of modifying the situation of the universe
prior to BBN. For example, in panel $(a)$, when $m=100$ GeV, $\sigma_0$ is 
increased from $4.91 \times 10^{-26}$ which corresponded to the standard 
expansion rate to $1.70 \times 10^{-25}$ for $n_D=8$ and to 
$3.85 \times 10^{-25}$ for $n_D=6$, 
respectively. If fixing $n_D$, the maximum value of
$\sigma_0$($\sigma_1$) examined by a certain aspect of the cosmological
perspective is determined by $\eta_{\rm max}$ which corresponded to the weakest-limit-value of BBN limit range, here
$\eta_{\max} = \left(Tr/{\rm 1MeV}\right)^{n_D-4}\,$ .
The increase of perturbative annihilation cross section is quite
sizable, that the extent of the enhancement could even exceeded to three or 
four orders of magnitude in the case which we discussed.
 Panels $(c)$ and $(d)$ are compatible with $(a)$ and $(b)$ through 
$ \sigma_0 = \pi \alpha^2/m^2$ and $\sigma_1 = 3 \pi \alpha^2/(8 m^2)$. Here 
the solid red lines are for $n_D=6$, the dashed blue lines are for $n_D=8$ and 
the dotted red lines are for the standard cosmology.
The maximum values of the mass for the fixed asymmetry factor are the same for 
various cases. For the same mass bound, stronger coupling
is needed for smaller $n_D$. We find the reason from Fig.\ref{yx}. The final 
abundance for $n_D=6$ is greater than the case of $n_D=8$. In order to satisfy
the observed value of DM abundance, one needs stronger coupling constant
$\alpha$. On the other hand, larger asymmetry factor leads the lower mass
bound. Compared to $s-$wave case, the $p-$wave annihilation cross section
requires stronger coupling as shown in panel $(d)$ in Fig.\ref{sigma-m}.  
\begin{figure}[h]
  \begin{center}
    \hspace*{-0.5cm} \includegraphics*[width=8cm]{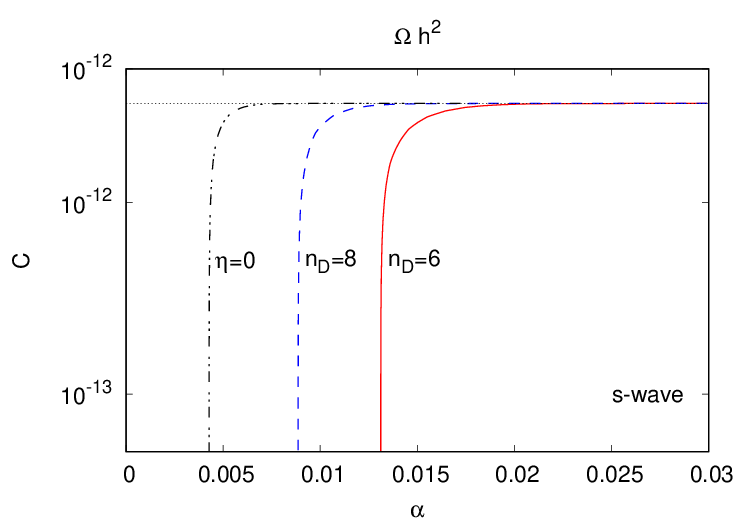}
    \put(-115,-12){(a)}
    \hspace*{-0.5cm} \includegraphics*[width=8cm]{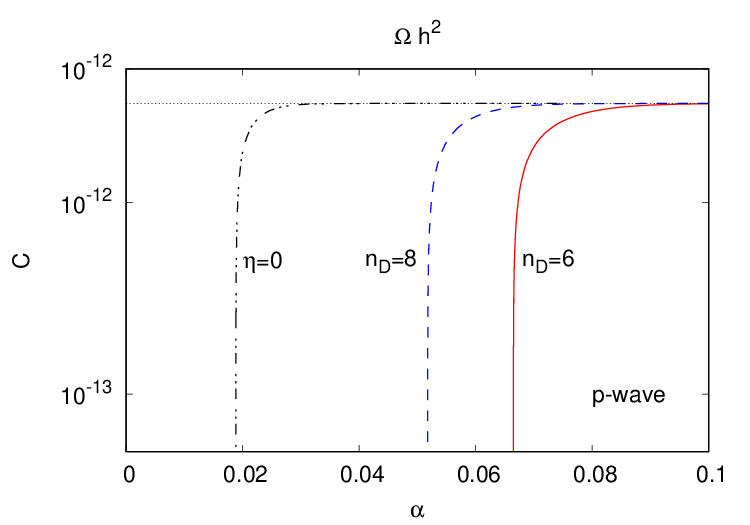}
    \put(-115,-12){(b)}
    \caption{\label{fig:c} 
    \footnotesize Contour plot of asymmetry factor $C$ 
    with the coupling strength $\alpha$ for $s-$wave ($p-$wave) 
   annihilation cross section when $\Omega_{\rm DM} h^2 = 0.120$. Here 
   $m = 130$ GeV for, $m_{\phi} = 0.25$ GeV,
    $g_{\chi} = 2$, $g_* = 90$.}
     \end{center}
\end{figure}

The contour plots of asymmetry factor $C$ and the coupling strength $\alpha$ 
are shown in Fig.\ref{fig:c}. We can see that when $\alpha$ is much larger, the
asymmetry factor $C$ required by the observed DM relic density does not depend 
on the expansion rate, as shown in the Fig.\ref{fig:c} where the three lines
corresponded to different cosmic expansion rates eventually converge into
one. For fixed $C$, the observed data requires the stronger coupling 
$\alpha$ when the expansion rate of the universe is enhanced. 
%In contrast with the
%$s-$wave annihilation, the $p-$wave annihilation needs stronger coupling. 

%

\section{Summary}
In this article, we investigated the relic abundance of asymmetric DM with Sommerfeld enhancement in non-standard cosmological scenarios. We first discussed the affect of the cosmology on the Boltzmann equation and subsequently introduced the additional energy density as the non-standard cosmological scenarios.  Analyzing the construction of relativistic and covariant Liouville operators, we find that the final form of the Boltzmann equation in non-standard cosmology differs from the form of that in standard cosmology only in the specific expression of cosmic expansion rate. 

In DM particle-anti-particle annihilation process, if the interaction length of the process is much larger than the Bohr radius, the incoming particle's wave functions can not be considered as free state at the origin, causing the
Sommerfeld effect. Then, we discussed the effect of Sommerfeld enhancement on
the density evolution of asymmetric DM when the expansion rate of universe is
enhanced. The mixed impact of non-standard cosmology and Sommerfeld effect lies in
the decoupling point from thermal equilibrium value and the freeze-out process of relic
density, as well as in the difference between DM particle and anti-particle.
At the early time of the evolution of asymmetric DM density, the
impact of Sommerfeld enhancement is not significant. However, the enhancement
of the expansion rate of the universe can be quite large. Taking both effects 
into account, the asymmetric DM decoupling point from the equilibrium is still advanced compared to the standard cosmology. Following the decrease of the temperature, the enhancement factor of the cosmic expansion rate is continuously decreased, while the
Sommerfeld factor is continuously increased. As a result, the
freeze-out process of relic density would last for a long time, which also  affect the final relic abundance. The two opposing effect on relic density and contrary trend of the strength development, results a significantly different and interesting asymmetric DM density evolution process, as shown in Fig.\ref{yx}.

Finally, we calculated the constraints on the parameter spaces, such as
perturbation annihilation cross section, the coupling strength, asymmetric DM 
mass and the asymmetry factor, to obtain the observed relic density of DM. Due 
to the introduction
of additional energy fields, the expansion rate of the
universe is enhanced. These non-standard cosmological models would predict the
more larger dark coupling strength and perturbative annihilation cross
section, through the observed DM relic abundance in the current universe. The
parameter spaces of DM particle would be broadened. 
For the non-standard cosmological models where the cosmic expansion rate is
reduced, the result is inverse. In the cases which we discussed, the broadening of the allowable range of the parameters of DM annihilation processes would give a wider region for DM detections and researchs. As a trade-off, the history of universe prior to BBN is changed. These findings demonstrate the implications of cosmological models on the interaction properties of DM particles. 

\section*{Acknowledgments}

The work is supported by the National Natural Science Foundation of China
(12463001).

\bibliographystyle{unsrt}
\bibliography{reference}

\begin{thebibliography}{10}

\bibitem{Planck:2018vyg}
N.~Aghanim et~al.
\newblock {Planck 2018 results. VI. Cosmological parameters}.
\newblock {\em Astron. Astrophys.}, 641:A6, 2020.
\newblock [Erratum: Astron.Astrophys. 652, C4 (2021)].

\bibitem{Hooper:2004dc}
Dan Hooper, John March-Russell, and Stephen~M. West.
\newblock {Asymmetric sneutrino dark matter and the Omega(b) / Omega(DM)
  puzzle}.
\newblock {\em Phys. Lett. B}, 605:228--236, 2005.

\bibitem{Nardi:2008ix}
Enrico Nardi, Francesco Sannino, and Alessandro Strumia.
\newblock {Decaying Dark Matter can explain the e+- excesses}.
\newblock {\em JCAP}, 01:043, 2009.

\bibitem{An:2009vq}
Haipeng An, Shao-Long Chen, Rabindra~N. Mohapatra, and Yue Zhang.
\newblock {Leptogenesis as a Common Origin for Matter and Dark Matter}.
\newblock {\em JHEP}, 03:124, 2010.

\bibitem{Iminniyaz:2011yp}
Hoernisa Iminniyaz, Manuel Drees, and Xuelei Chen.
\newblock {Relic Abundance of Asymmetric Dark Matter}.
\newblock {\em JCAP}, 07:003, 2011.

\bibitem{Cohen:2009fz}
Timothy Cohen and Kathryn~M. Zurek.
\newblock {Leptophilic Dark Matter from the Lepton Asymmetry}.
\newblock {\em Phys. Rev. Lett.}, 104:101301, 2010.

\bibitem{Kaplan:2009ag}
David~E. Kaplan, Markus~A. Luty, and Kathryn~M. Zurek.
\newblock {Asymmetric Dark Matter}.
\newblock {\em Phys. Rev. D}, 79:115016, 2009.

\bibitem{Cohen:2010kn}
Timothy Cohen, Daniel~J. Phalen, Aaron Pierce, and Kathryn~M. Zurek.
\newblock {Asymmetric Dark Matter from a GeV Hidden Sector}.
\newblock {\em Phys. Rev. D}, 82:056001, 2010.

\bibitem{Shelton:2010ta}
Jessie Shelton and Kathryn~M. Zurek.
\newblock {Darkogenesis: A baryon asymmetry from the dark matter sector}.
\newblock {\em Phys. Rev. D}, 82:123512, 2010.

\bibitem{Kang:2011cni}
Zhaofeng Kang, Jinmian Li, Tianjun Li, Tao Liu, and Jin~Min Yang.
\newblock {The maximal $U(1)_L$ inverse seesaw from $d=5$ operator and
  oscillating asymmetric Sneutrino dark matter}.
\newblock {\em Eur. Phys. J. C}, 76(5):270, 2016.

\bibitem{Ellwanger:2012yg}
Ulrich Ellwanger and Pantelis Mitropoulos.
\newblock {Upper Bounds on Asymmetric Dark Matter Self Annihilation Cross
  Sections}.
\newblock {\em JCAP}, 07:024, 2012.

\bibitem{Petraki:2013wwa}
Kalliopi Petraki and Raymond~R. Volkas.
\newblock {Review of asymmetric dark matter}.
\newblock {\em Int. J. Mod. Phys. A}, 28:1330028, 2013.

\bibitem{Boucenna:2013wba}
S.~M. Boucenna and S.~Morisi.
\newblock {Theories relating baryon asymmetry and dark matter: A mini review}.
\newblock {\em Front. in Phys.}, 1:33, 2014.

\bibitem{Zurek:2013wia}
Kathryn~M. Zurek.
\newblock {Asymmetric Dark Matter: Theories, Signatures, and Constraints}.
\newblock {\em Phys. Rept.}, 537:91--121, 2014.

\bibitem{Mahapatra:2023dbr}
Satyabrata Mahapatra, Partha~Kumar Paul, Narendra Sahu, and Prashant Shukla.
\newblock {Asymmetric long-lived dark matter and leptogenesis from the type-III
  seesaw framework}.
\newblock {\em Phys. Rev. D}, 111(1):015043, 2025.

\bibitem{Borah:2024wos}
Debasish Borah, Satyabrata Mahapatra, Partha~Kumar Paul, Narendra Sahu, and
  Prashant Shukla.
\newblock {Asymmetric self-interacting dark matter with a canonical seesaw
  model}.
\newblock {\em Phys. Rev. D}, 110(3):035033, 2024.

\bibitem{Baldes:2017gzw}
Iason Baldes and Kalliopi Petraki.
\newblock {Asymmetric thermal-relic dark matter: Sommerfeld-enhanced
  freeze-out, annihilation signals and unitarity bounds}.
\newblock {\em JCAP}, 09:028, 2017.

\bibitem{Bollig:2024ipe}
Julian Bollig.
\newblock {\em {The impact of non-perturbative effects in dark matter
  production and detection}}.
\newblock PhD thesis, Freiburg U., 2024.

\bibitem{Arkani-Hamed:2008hhe}
Nima Arkani-Hamed, Douglas~P. Finkbeiner, Tracy~R. Slatyer, and Neal Weiner.
\newblock {A Theory of Dark Matter}.
\newblock {\em Phys. Rev. D}, 79:015014, 2009.

\bibitem{Sommerfeld:1931qaf}
A.~Sommerfeld.
\newblock {\"Uber die Beugung und Bremsung der Elektronen}.
\newblock {\em Annalen Phys.}, 403(3):257--330, 1931.

\bibitem{Qiu:2024iyo}
Sujuan Qiu, Hoernisa Iminniyaz, and Wensheng Huo.
\newblock {Asymmetric dark matter and Sommerfeld enhancement}.
\newblock {\em Commun. Theor. Phys.}, 76(8):085403, 2024.

\bibitem{Iengo:2009ni}
Roberto Iengo.
\newblock {Sommerfeld enhancement: General results from field theory diagrams}.
\newblock {\em JHEP}, 05:024, 2009.

\bibitem{Slatyer:2009vg}
Tracy~R. Slatyer.
\newblock {The Sommerfeld enhancement for dark matter with an excited state}.
\newblock {\em JCAP}, 02:028, 2010.

\bibitem{Petraki:2015hla}
Kalliopi Petraki, Marieke Postma, and Michael Wiechers.
\newblock {Dark-matter bound states from Feynman diagrams}.
\newblock {\em JHEP}, 06:128, 2015.

\bibitem{vonHarling:2014kha}
Benedict von Harling and Kalliopi Petraki.
\newblock {Bound-state formation for thermal relic dark matter and unitarity}.
\newblock {\em JCAP}, 12:033, 2014.

\bibitem{Petraki:2016cnz}
Kalliopi Petraki, Marieke Postma, and Jordy de~Vries.
\newblock {Radiative bound-state-formation cross-sections for dark matter
  interacting via a Yukawa potential}.
\newblock {\em JHEP}, 04:077, 2017.

\bibitem{Petraki:2014uza}
Kalliopi Petraki, Lauren Pearce, and Alexander Kusenko.
\newblock {Self-interacting asymmetric dark matter coupled to a light massive
  dark photon}.
\newblock {\em JCAP}, 07:039, 2014.

\bibitem{Kamada:2020buc}
Ayuki Kamada, Hee~Jung Kim, and Takumi Kuwahara.
\newblock {Maximally self-interacting dark matter: models and predictions}.
\newblock {\em JHEP}, 12:202, 2020.

\bibitem{Cassel:2009wt}
S.~Cassel.
\newblock {Sommerfeld factor for arbitrary partial wave processes}.
\newblock {\em J. Phys. G}, 37:105009, 2010.

\bibitem{Duerr:2018mbd}
Michael Duerr, Kai Schmidt-Hoberg, and Sebastian Wild.
\newblock {Self-interacting dark matter with a stable vector mediator}.
\newblock {\em JCAP}, 09:033, 2018.

\bibitem{Feng:2010zp}
Jonathan~L. Feng, Manoj Kaplinghat, and Hai-Bo Yu.
\newblock {Sommerfeld Enhancements for Thermal Relic Dark Matter}.
\newblock {\em Phys. Rev. D}, 82:083525, 2010.

\bibitem{Cirelli:2016rnw}
Marco Cirelli, Paolo Panci, Kalliopi Petraki, Filippo Sala, and Marco Taoso.
\newblock {Dark Matter's secret liaisons: phenomenology of a dark U(1) sector
  with bound states}.
\newblock {\em JCAP}, 05:036, 2017.

\bibitem{Feng:2009mn}
Jonathan~L. Feng, Manoj Kaplinghat, Huitzu Tu, and Hai-Bo Yu.
\newblock {Hidden Charged Dark Matter}.
\newblock {\em JCAP}, 07:004, 2009.

\bibitem{Agrawal:2017rvu}
Prateek Agrawal, Francis-Yan Cyr-Racine, Lisa Randall, and Jakub Scholtz.
\newblock {Dark Catalysis}.
\newblock {\em JCAP}, 08:021, 2017.

\bibitem{Agrawal:2016quu}
Prateek Agrawal, Francis-Yan Cyr-Racine, Lisa Randall, and Jakub Scholtz.
\newblock {Make Dark Matter Charged Again}.
\newblock {\em JCAP}, 05:022, 2017.

\bibitem{Chen:2023rrl}
Zien Chen, Kairui Ye, and Mengchao Zhang.
\newblock {Asymmetric dark matter with a spontaneously broken U(1)':
  Self-interaction and gravitational waves}.
\newblock {\em Phys. Rev. D}, 107(9):095027, 2023.

\bibitem{Joyce:1996cp}
Michael Joyce.
\newblock {Electroweak Baryogenesis and the Expansion Rate of the Universe}.
\newblock {\em Phys. Rev. D}, 55:1875--1878, 1997.

\bibitem{Iminniyaz:2016iom}
Hoernisa Iminniyaz.
\newblock {Asymmetric Dark Matter in the Shear--dominated Universe}.
\newblock {\em Phys. Lett. B}, 765:6--10, 2017.

\bibitem{Drees:2007kk}
Manuel Drees, Hoernisa Iminniyaz, and Mitsuru Kakizaki.
\newblock {Constraints on the very early universe from thermal WIMP dark
  matter}.
\newblock {\em Phys. Rev. D}, 76:103524, 2007.

\bibitem{Drees:2006vh}
Manuel Drees, Hoernisa Iminniyaz, and Mitsuru Kakizaki.
\newblock {Abundance of cosmological relics in low-temperature scenarios}.
\newblock {\em Phys. Rev. D}, 73:123502, 2006.

\bibitem{Salati:2002md}
Pierre Salati.
\newblock {Quintessence and the relic density of neutralinos}.
\newblock {\em Phys. Lett. B}, 571:121--131, 2003.

\bibitem{BINETRUY2000285}
Pierre Binétruy, Cédric Deffayet, Ulrich Ellwanger, and David Langlois.
\newblock Brane cosmological evolution in a bulk with cosmological constant.
\newblock {\em Physics Letters B}, 477(1):285--291, 2000.

\bibitem{Iminniyaz:2018das}
Hoernisa Iminniyaz, Burhan Salai, and Guoliang Lv.
\newblock {Relic Density of Asymmetric Dark Matter in Modified Cosmological
  Scenarios}.
\newblock {\em Commun. Theor. Phys.}, 70(5):602, 2018.

\bibitem{Binetruy:1999ut}
Pierre Binetruy, Cedric Deffayet, and David Langlois.
\newblock {Nonconventional cosmology from a brane universe}.
\newblock {\em Nucl. Phys. B}, 565:269--287, 2000.

\bibitem{DEramo:2017gpl}
Francesco D'Eramo, Nicolas Fernandez, and Stefano Profumo.
\newblock {When the Universe Expands Too Fast: Relentless Dark Matter}.
\newblock {\em JCAP}, 05:012, 2017.

\bibitem{Langlois:2002bb}
David Langlois.
\newblock {Brane cosmology: An Introduction}.
\newblock {\em Prog. Theor. Phys. Suppl.}, 148:181--212, 2003.

\bibitem{Guo:2009nt}
Wan-Lei Guo and Xin Zhang.
\newblock {Constraints on Dark Matter Annihilation Cross Section in Scenarios
  of Brane-World and Quintessence}.
\newblock {\em Phys. Rev. D}, 79:115023, 2009.

\bibitem{Cicoli:2023opf}
Michele Cicoli, Joseph~P. Conlon, Anshuman Maharana, Susha Parameswaran,
  Fernando Quevedo, and Ivonne Zavala.
\newblock {String cosmology: From the early universe to today}.
\newblock {\em Phys. Rept.}, 1059:1--155, 2024.

\bibitem{Catena:2009tm}
R.~Catena, N.~Fornengo, M.~Pato, L.~Pieri, and A.~Masiero.
\newblock {Thermal Relics in Modified Cosmologies: Bounds on Evolution
  Histories of the Early Universe and Cosmological Boosts for PAMELA}.
\newblock {\em Phys. Rev. D}, 81:123522, 2010.

\bibitem{Hertzberg:2024uqy}
Mark~P. Hertzberg and Abraham Loeb.
\newblock {Constraints on an anisotropic universe}.
\newblock {\em Phys. Rev. D}, 109(8):083538, 2024.

\bibitem{Gron:2024vmf}
\O{}yvind~G. Gr\o{}n.
\newblock {Anisotropic Generalization of the \ensuremath{\Lambda}CDM Universe
  Model with Application to the Hubble Tension}.
\newblock {\em Symmetry}, 16(5):564, 2024.

\bibitem{Okada:2004nc}
Nobuchika Okada and Osamu Seto.
\newblock {Relic density of dark matter in brane world cosmology}.
\newblock {\em Phys. Rev. D}, 70:083531, 2004.

\bibitem{AbouElDahab:2006glf}
E.~Abou El~Dahab and S.~Khalil.
\newblock {Cold dark matter in brane cosmology scenario}.
\newblock {\em JHEP}, 09:042, 2006.

\bibitem{Schelke:2006eg}
Mia Schelke, Riccardo Catena, Nicolao Fornengo, Antonio Masiero, and Massimo
  Pietroni.
\newblock {Constraining pre Big-Bang-Nucleosynthesis Expansion using Cosmic
  Antiprotons}.
\newblock {\em Phys. Rev. D}, 74:083505, 2006.

\bibitem{Giare:2024akf}
William Giar\`e.
\newblock {Inflation, the Hubble tension, and early dark energy: An alternative
  overview}.
\newblock {\em Phys. Rev. D}, 109(12):123545, 2024.

\bibitem{Chanda:2021tzi}
Prolay Chanda and James Unwin.
\newblock {Decoupling of asymmetric dark matter during an early matter
  dominated era}.
\newblock {\em JCAP}, 06:032, 2021.

\bibitem{CBLiang}
Canbin Liang and Bin Zhou.
\newblock {\em Differential Geometry and General Relativity}.
\newblock Springer Singapore, 2023.

\bibitem{reviewBE}
Carlo Cercignani and Gilberto~Medeiros Kremer.
\newblock {\em The relativistic boltzmann equation: theory and applications}.
\newblock Birkhäuser Basel, 2002.

\bibitem{Weinberg:2008zzc}
Steven Weinberg.
\newblock {\em {Cosmology}}.
\newblock 2008.

\bibitem{Kolb:1988aj}
Edward~W. Kolb and Michael~S. Turner.
\newblock {\em {The Early Universe}}.
\newblock 1988.

\bibitem{Scherrer:1985zt}
Robert~J. Scherrer and Michael~S. Turner.
\newblock {On the Relic, Cosmic Abundance of Stable Weakly Interacting Massive
  Particles}.
\newblock {\em Phys. Rev. D}, 33:1585, 1986.
\newblock [Erratum: Phys.Rev.D 34, 3263 (1986)].

\end{thebibliography}

\end{document}